%% file: acl.tex
\pdfoutput=1

\documentclass[11pt]{article}
\usepackage[table,dvipsnames]{xcolor}
\usepackage[preprint]{acl}

\usepackage{times}
\usepackage{latexsym}
\usepackage{booktabs}
\usepackage{multirow}
\usepackage{enumitem}
\usepackage[T1]{fontenc}
\usepackage{graphicx}
\usepackage[utf8]{inputenc}
\usepackage{microtype}
\usepackage{caption}
\usepackage{inconsolata}
\usepackage{amsmath}
\usepackage{hyperref}
\usepackage{cleveref}
\usepackage{bm}
\usepackage{amsfonts}
\usepackage{enumitem}
\definecolor{lightgray}{RGB}{236,236,236} 
\title{Exploring the Best Practices of Query Expansion with Large Language Models }

\author{
   Le Zhang$^{1,2, \ast}$, Yihong Wu$^{2, \ast}$, Qian Yang$^{1,2}$,  Jian-Yun Nie$^2$\\
  $^1$ Mila - Québec AI Institute \\
  $^2$ Université de Montréal
}

\definecolor{darkgreen}{rgb}{0.0, 0.5, 0.0}

\begin{document}
\maketitle
\input{chapter/abstract}

\input{chapter/introduction}

\input{chapter/related_work}

\input{chapter/method}

\input{chapter/exp}
\input{chapter/analysis}
\input{chapter/conclusion}

\section*{Limitation}
The proposed method offers substantial improvements in information retrieval performance. Nonetheless, a shared limitation across query expansion approaches, ours included, is the increased inference time due to the generation of extra passages with LLMs. Retrieval-Augmented Generation presents a potential solution by interleaving the retrieval and generation processes, potentially mitigating this issue, which requires further exploration.

\bibliography{acl}

%
\clearpage
\end{document}

%% file: chapter/abstract.tex
\begin{abstract}

Large Language Models (LLMs) are foundational in language technologies, particularly in information retrieval (IR). Previous studies have utilized LLMs for query expansion, achieving notable improvements in IR. In this paper, we thoroughly explore the best practice of leveraging LLMs for query expansion. To this end, we introduce a training-free, straightforward yet effective framework called Multi-Text Generation Integration (\textsc{MuGI}). It leverages LLMs to generate multiple pseudo-references, integrating them with queries to enhance both sparse and dense retrievers. Our empirical findings reveal that: (1) Increasing the number of samples from LLMs benefits IR systems; (2) A balance between the query and pseudo-documents, and an effective integration strategy, is critical for high performance; (3) Contextual information from LLMs is essential, even boost a 23M model to outperform a 7B baseline model; (4) Pseudo relevance feedback can further calibrate queries for improved performance; and (5) Query expansion is widely applicable and versatile, consistently enhancing models ranging from 23M to 7B parameters. Our code and all generated references are made available at \url{https://github.com/lezhang7/Retrieval_MuGI}

\end{abstract}

%% file: chapter/introduction.tex
\section{Introduction}

Information retrieval (IR) is crucial for extracting relevant documents from large databases, serving as a key component in search engines, dialogue systems~\cite{Yuan2019MultihopSN,Qian2021LearningIU}, question-answering platforms~\cite{Qu2020RocketQAAO,zhang-etal-2023-moqagpt,Yang_SKURG_23}, recommendation systems~\cite{Fan2023RecommenderSI,Zhang2023RecommendationAI}, and Retrieval Augmented Generation (RAG)~\cite{lewis2021retrievalaugmented,izacard2020leveraging,zhang2022retgen, Liu_LlamaIndex_2022}.


Query expansion, a key technique for enhancing information retrieval (IR) efficacy \cite{abdul2004umass, Robertson1976RelevanceWO, Salton1971TheSR}, traditionally employs Pseudo-Relevance Feedback (PRF) \cite{li2022pseudo, lavrenko2017relevance} from initial retrieval results. However, its effectiveness is constrained by the quality of these results. Recently, Large Language Models (LLMs), such as ChatGPT, have demonstrated exceptional capabilities in language understanding, knowledge storage, and reasoning~\cite{gpt3,Touvron2023LLaMAOA}. Motivated by these advancements, some studies have explored leveraging LLMs for zero-shot query expansion~\cite{ma2023zero,hyde,wang2023query2doc}. While these methods have shown empirical effectiveness, they also present certain limitations.

LameR~\cite{lamer} generates potential answers by utilizing LLMs to rewrite BM25 candidates for expansion. However, its performance is highly dependent on the quality of the initial retrieval. Both HyDE~\cite{hyde} and query2doc~\cite{wang2023query2doc} leverage the knowledge stored in LLMs. While HyDE demonstrates effective performance with contriver, it performs poorly with lexical-based retrievers~\cite{lamer}. Conversely, query2doc is effective with both sparse and dense retrieval methods, but strong rankers may not benefit as much as weaker ones~\cite{li2024can, Weller2023WhenDG}. Moreover, the integration and balance between pseudo references and queries are under-explored in these studies.

To address these limitations, we explore best practices for utilizing query expansion with LLMs for information retrieval. In this paper, we delve into several specific research questions: \textbf{RQ1}: Are multiple pseudo-references more beneficial than a single one? \textbf{RQ2}: Is there a universal query expansion method that effectively serves both lexical-based and neural-based retrievers, applicable to both weak and strong models without prior constraints? \textbf{RQ3}: How can the query and pseudo-references be balanced for lexical-based retrievers? \textbf{RQ4}: What is the most effective method for integrating multiple pseudo-references with a query in dense retrievers?

We introduce a framework named \textbf{Mu}lti-\textbf{T}ext \textbf{G}eneration \textbf{I}ntegration (\textbf{\textsc{MuGI}}) to address these key questions. \textsc{MuGI} employs a zero-shot approach to generate multiple pseudo-references from LLMs, integrating them with queries to enhance IR efficiency. Our empirical experiments demonstrate that:
(1) Increasing the number of samples from LLMs benefits IR systems.
(2) \textsc{MuGI} demonstrates versatility and effectiveness across both lexical and dense retrievers and models of various sizes. Remarkably, it enables a 23M-parameter dense retriever to outperform a larger 7B baseline.
(3) \textsc{MuGI} proposes an adaptive reweighting strategy that considers the lengths of both the pseudo-references and the query, critically improving the performance of lexical retrievers.
(4) \textsc{MuGI} investigates different integration strategies and proposes contextualized pooling, which has been overlooked in previous methods.
Additionally, drawing inspiration from the Rocchio algorithm~\cite{schutze2008introduction}, \textsc{MuGI} implements a calibration module that leverages pseudo relevance feedback to further enhance IR performance. Notably, using ChatGPT4, \textsc{MuGI} significantly enhances BM25 performance, with an 18\% improvement on the TREC DL dataset and 7.5\% on BEIR, and boosts dense retrievers by over 7\% on TREC DL and 4\% on BEIR.

%% file: chapter/related_work.tex
\section{Related Work}
\paragraph{Information Retrieval} 
focuses on the efficient and effective retrieval of information in response to user queries. Best Matching 25 (BM25)~\cite{BM25} advances beyond earlier probabilistic models by incorporating document length normalization and non-linear term frequency scaling, thereby enhancing the alignment of queries with documents. Dense retrievers such as DPR~\cite{dpr} employ deep neural networks to identify semantic relationships between queries and documents by measuring the cosine similarity of their text embeddings.

Existing efficient IR systems typically use a retrieval \& rerank pipeline~\cite{nogueira2020passage, dpr, Guo_2022, reimers-2019-sentence-bert}: Initially, a retrieval mechanism, such as BM25 or a bi-encoder, identifies a broad set of potentially relevant documents. Subsequently, a stronger ranker, usually a cross-encoder, meticulously scores the relevance of these documents, enhancing the precision of the final results.  

\paragraph{LLMs for IR} The use of LLMs in IR falls into two primary categories~\cite{zhu2023large}: fine-tuning LLMs as retrieval models and employing them for zero-shot IR. This paper concentrates on zero-shot IR, where typical approaches involve leveraging the reasoning capabilities of LLMs for direct document ranking~\cite{rankgpt,ma2023zero} or relevance assessment~\cite{UPR}. While effective, these methods are limited by LLMs' input length constraints, making them better suited for the rerank phase.

Another line of research focuses on using LLMs to synthesize additional high-quality training datasets to improve existing models~\cite{bonifacio2022inpars,izacard2021unsupervised,E5mistral7b,jeronymo2023inpars}.
Other works, such as HyDE~\cite{hyde}, query2doc~\cite{wang2023query2doc}, and LameR~\cite{lamer}, explore query expansion. They leverage LLMs to create pseudo-references or potential answers, enhancing queries for better retrieval outcomes. 

MuGI is a query expansion framework that leverages LLMs to enhance queries. Unlike previous works, which are limited by inherent constraints, MuGI offers broader applicability and versatility as it seamlessly integrates with both lexical and dense retrievers.
By utilizing and intergrating a wealth of contextualized information from multiple references, MuGI surpasses existing techniques in both in-domain and out-of-distribution evaluations by more effectively capturing essential keywords and enriching the background context.

%% file: chapter/method.tex
\section{Method}

We begin by discussing IR preliminaries and introducing our MuGI framework, which is designed to address the questions outlined earlier.

\subsection{Preliminaries}

\paragraph{Non-parametric Lexical-based Methods}
BM25 is a fundamental non-parametric lexical method that calculates document relevance using term frequency (TF) and inverse document frequency (IDF) as:. 
\begin{equation}
\begin{aligned}
\small
\sum_{i=1}^{n} \frac{\text{IDF}(q_i)\text{TF}(q_i, D)  (k_1 + 1)}{\text{TF}(q_i, D) + k_1 (1 - b + b \frac{|D|}{\text{avgdl}})}
\end{aligned}
\end{equation}
where $q_i$ are query terms, $\text{TF}(q_i, D)$ is term frequency, $\text{IDF}(q_i)$ is inverse document frequency, $|D|$ is document length, $\text{avgdl}$ is average document length, and $k_1$ and $b$ are tunable parameters.

\paragraph{Neural Dense Retrieval Methods}
Dense retrieval leverages deep learning to identify semantic similarities between queries and documents by encoding them into high-dimensional embeddings, typically measured by \cite{huang2013learning}:

\newcommand{\f}{\bm{f}}
\begin{equation}
\text{Sim}(q, D) = \frac{\f(q)^\top \f(D)}{\|\f(q)\| \|\f(D)\|}
\end{equation}
where $\f(\cdot)$ maps text to embedding space $\mathbb{R}^d$.
BM25 is fast and generalizes well, suited for sparase retrieval, while dense retrieval excels at capturing semantic connections but is slower and less generalized due to neural network dependency.

\subsection{Multi-Text Generation Integration}

\begin{figure*}[!th]
    \centering
    \includegraphics[width=\textwidth]{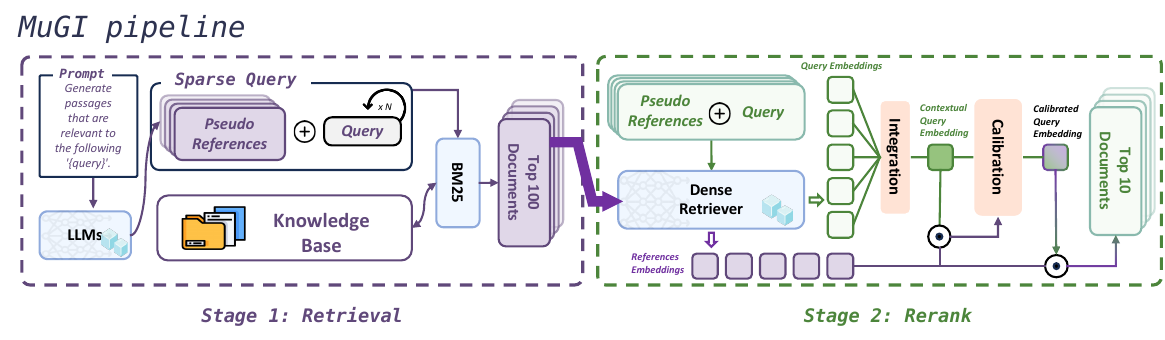}
    \caption{\textbf{Method overview of MuGI.} Left part is initial retrieval using BM25 for initial retrieval, right part indicates re-rank output from first stage using a dense retriever.}
    \label{fig:method}
    \vspace{-5mm}
\end{figure*}

Recognizing that both lexical-based and dense retrieval methods depend on a certain degree of information overlap between the query and document, we introduce the \textbf{Multi-Text Generation Integration} (\textsc{MuGI}) method.
This approach aims to augment the query's information content by leveraging multiple samplings from LLMs.
\textsc{MuGI} enriches queries with additional background information and broadens the keyword vocabulary to encompass out-of-domain terms, thereby bridging the semantic gap between queries and documents on both lexical-based and dense retrievers.
Figure \ref{fig:method} provides an illustrative overview of \textsc{MuGI}.

\begin{figure}[th]
    \centering
    \includegraphics[width=\linewidth]{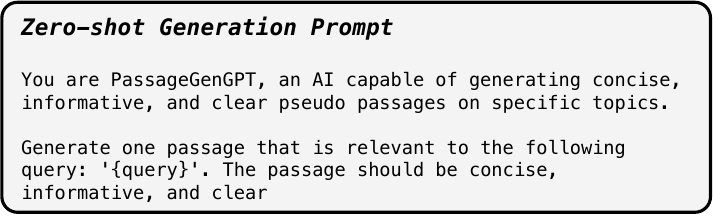}
    \caption{Zero-Shot Prompting for Relevant Passage Generation: It emphasizes generating contextually relevant content to enhance background knowledge density for multiple-text integration.}
    \label{fig:zsprompt}
    \vspace{-3mm}
\end{figure}

Upon receiving a query \textit{q}, \textsc{MuGI} initially applies a zero-shot prompt (see \cref{fig:zsprompt}) technique to generate a set of pseudo-references, denoted as $\mathcal{R} = \{r_1, r_2, r_3, ..., r_n\}$, which are then integrated with query for subsequent IR operations. We have explored different methods for BM25 and dense retrievers.

\subsubsection{\textsc{MuGI} for BM25} This component evaluates relevance by analyzing lexical overlaps between the query and references. Given the longer lengths of documents compared to queries and BM25's sensitivity to word frequency, achieving a careful balance to ensure the appropriate influence of each element in text is crucial. The variation in the lengths of queries and passages makes the constant repetition of query used in previous studies, which typically handles single pseudo-references, ineffective \citep{wang2023query2doc,lamer}, particularly when dealing with multiple references.

To address this issue, we implement an adaptive reweighting strategy that adjusts according to the length of the pseudo-references. This adjustment is governed by a factor $\beta$, as illustrated by the following equation:
\begin{align}
\lambda = \left\lfloor \frac{\mathrm{len}(r_1)+\mathrm{len}(r_2)+\ldots+\mathrm{len}(r_n))}{\mathrm{len}(q) \cdot \beta} \right\rfloor \label{eq}
\end{align}

Since BM25 does not account for word order, we enhance the query by repeating query $\lambda$ times and concatenating it with all pseudo-references:
\begin{align}
q_{\text{sparse}} = \text{concat}({q}*\lambda, r_1, r_2, r_3..., r_n)
\end{align}
This enhanced query is then processed by BM25 to produce the ranking results $I_\text{bm25}$.

\subsubsection{\textsc{MuGI} for Dense Retriever}
\textsc{MuGI} also enhances dense retrievers, specifically bi-encoders. In this section, we discuss how to integrate pseudo-references with queries and refine query using pseudo positive/negative reference feedback.
\paragraph{Integration} We present two approaches to integrate queries with pseudo-references to obtain a contextualized query embedding.

\begin{enumerate}[label=\Roman*., itemsep=1pt, topsep=0pt, left=0pt]
    \item \textbf{Concatenation} has been commonly used in prior studies \cite{lamer, wang2023query2doc}, where the query is simply concatenated with all references as in BM25:
\begin{align}
q_{\text{cat}} = \text{concat}(q, r_1, r_2, ..., r_n)
\end{align}
This enhanced query is then processed by the dense retriever $\f$ to produce embeddings, i.e., $\bm{e}_\text{cat} = \bm{f}(q_\text{cat})$.
However, as the number and length of references increase, the typical input length limitation of 512 tokens can hinder the integration process. Consequently, only one to two passages can be incorporated into $q_{cat}$.
    \item \textbf{Feature Pooling} addresses the model's input length limitations, particularly when multiple references are involved. A straightforward method is to average the embeddings in the feature space, as demonstrated by HyDE \cite{hyde}:
  \begin{align}
    \bm{e}_{\text{mean-pool}} = \frac{\f(q)+\sum_{i} \f(r_i)}{n+1}
 \end{align}
    Empirically, we use another variant termed contex-pool:
      \begin{align}
    \bm{e}_{\text{contex-pool}} = \frac{\sum_{i} \f(\text{concat}(q,r_i)}{n}
 \end{align}
\end{enumerate}

We then calculate the similarity between the query and all documents, ranking them accordingly. We denote these rankings as $I_\text{pre}$.

\paragraph{Calibration}
The Rocchio algorithm refines the query vector using relevance feedback as follows \cite{schutze2008introduction}:
\begin{equation}
    \bm{e}'_q = a\bm{e}_q + \frac{b}{|\mathcal{D}_+|}\sum_{\bm{d}_i\in \mathcal{D}_+ }\bm{d}_i - \frac{c}{|\mathcal{D}_-|}\sum_{\bm{d}_j\in \mathcal{D}_-}\bm{d}_j,
\end{equation}
where $\bm{e}'_q$ is the calibrated query vector, $\bm{e}_q$ is the original query vector, $\bm{d}_i, \bm{d}_j$ are document vectors, $\mathcal{D}_+$/$\mathcal{D}_-$ is the set of positive/negative documents, and $a,b,c$ are weights.
Given that the Rocchio algorithm is tailored for bag-of-words methods and not optimized for neural models, we introduce an adapted post-processing operation for neural-based dense retriever:

We construct a negative feedback set $\mathcal{N}$ from the last $n$ documents in the BM25 results, and a positive feedback set $\mathcal{R}+$ from all generated references and the K-reciprocal documents, i.e the intersection of the top-K documents in $I_\text{bm25}$ and $I_\text{pre}$. The adjusted calibration process is:
\begin{equation}
    \bm{e}'_q = \frac{1}{W}(\sum_{r\in \mathcal{R}_+}\f(\text{concat}(q, r)) - \alpha\sum_{n'\in\mathcal{N}}\f(n')),
\end{equation}
where $W=|\mathcal{R}|+|\mathcal{N}|$ is the total number of positive and negative feedback, $\alpha$ is the weight factor. The calibrated query is then used for ranking, producing the final result $I_\text{post}$. Note that calibration does not add extra computational overhead, as it operates directly on the output embedding.

\subsection{MuGI Pipeline} \label{sec
} Our MuGI framework is designed to enhance lexical-based methods and neural dense retrievers, particularly focusing on bi-encoders. We introduce a fast and high-quality retrieval and rerank pipeline termed \textsc{MuGI} Pipeline (see \cref{fig:method}) that begins by retrieving the top 100 references using BM25, and then reranks them using an enhanced query embedding from a \textbf{ bi-encoder model} for two primary reasons: 1) We aim to explore the impacts of query expansion on general dense embedding similarity models, specifically bi-encoders, as opposed to cross-encoders which require additional training and are more computationally costly. 2) Employing large bi-encoders for retrieval from large database, particularly those based on LLMs with over 7B parameters, is computationally intensive. 
Consequently, we limit the search scope of the bi-encoder to the top 100 references from BM25 and utilize these rerank results to demonstrate the effectiveness of the MuGI framework across bi-encoders of various sizes.

%% file: chapter/exp.tex
\input{tables/retrieval}
\section{Experiments}

In this section, We use \textsc{MuGI} to answer four questions brought in introduction section with comprehensive experiments.
\label{sec:exp}
\subsection{Setup}

\paragraph{Implementation Details} 
\label{sec:detail}
We employ ChatGPT, and Qwen \cite{qwen} to generate pseudo-references. For BM25 searches, we use the Pyserini toolkit \cite{Lin_etal_SIGIR2021_Pyserini} with default settings. We  select multiple dense bi-encoders based on their reranking performance in the MTEB leaderboard \cite{muennighoff2022mteb}, including: \textsc{all-MiniLM-L6-v2}, \textsc{ALL-MPNET-BASE-V2}, \textsc{BGE-Large-EN-v1.5} \cite{bge_embedding}, \textsc{Ember-v1}, and \textsc{E5-Mistral-instruct} \cite{E5mistral7b}, which vary in size from 23M to 7B parameters. We also incorporate strong cross-encoders, including MonoT5 \cite{monot5} and RankLLaMA \cite{ma2023fine}.

In our experiments, unless specified otherwise, \textsc{MuGI} uses a $\beta=4$ for BM25 and a context feature pool with a calibration factor of $\alpha=0.2$ for dense retrieval. Each retrieval process involves using $n=5$ generated pseudo-references from ChatGPT-4-turo-1106. All experiments are conducted with 1 NVIDIA A100 GPU.


\paragraph{Evaluation} Following \cite{wang2023query2doc}, we adopt nDCG@10 as the metric and evaluate \textsc{MuGI} on the TREC DL19~\cite{dl19} and DL20~\cite{dl20} datasets for in-domain analysis, and on nine low-resource datasets from BEIR~\cite{thakur2021beir} for out-of-distribution (OOD) evaluation. In benchmark dataset, we evaluate BM25 with sparse retrieval task and evaluate dense retriever with rerank task on the top-100 passages retrieved by BM25 following~\citet{rankgpt}.

\subsection{\textsc{MuGI} Result}

\paragraph{Applicability} is crucial for query expansion methods. Previous approaches like HyDE targeted only dense retrievers, while query2doc yields inconsistent results across models of varying capacities \cite{li2024can, Weller2023WhenDG}. Conversely, our \textsc{MuGI} framework enhances both BM25 and dense retrievers across all evaluations, as demonstrated in \cref{table:retrieval}
 and \cref{tab:rerank}. Unlike query2doc, \textsc{MuGI} consistently enhances models across a wide range of sizes, from 23M to 7B parameters, demonstrating its broad applicability.

\input{tables/rerank}

\paragraph{Sparse Retrievers}

For sparse retrieval evaluation, we include strong baselines such as HyDE~\cite{hyde}, Query2Doc~\cite{wang2023query2doc}, and LameR~\cite{lamer}, alongside compact dense retrievers like ANCE \cite{ance} and DPR \cite{dpr}. The results, presented in Table~\ref{table:retrieval}, show that \textsc{MuGI} outperforms all baseline models and existing query expansion techniques on the TREC DL dataset. Specifically, it boosts BM25 performance by 19.8\% and 16.4\% in nDCG@10 for the TREC DL 19/20 datasets, respectively.

\textsc{MuGI} also shows substantial improvements on the BEIR dataset, where queries are typically short and ambiguous. By integrating multiple pseudo-references, \textsc{MuGI} effectively enriches the context, leading to a 7.6\% enhancement over the baseline BM25 and outperforming other query expansion strategies across all tested datasets.

\paragraph{Dense Retrievers} Rerank involves reordering passages initially retrieved by sparse retrieval using advanced neural models. Our findings, shown in \cref{tab:rerank}, underscore the \textsc{MuGI} pipeline's substantial enhancement of dense retrievers' performance. The \textit{BM25 rerank} baseline reranks the top 100 references from BM25, while \textit{query2doc} applies the query2doc\footnote{Only the TREC DL dataset is released for query2doc.} method in both sparse retrieval and reranking phases. The \textit{MuGI pipeline}, outlined in \cref{fig:method}, integrates calibration into the reranking process.

Integrating \textsc{MuGI} into the reranking process consistently boosts performance, outperforming query2doc in all scenarios. This integration leads to over a 7\% improvement in \textbf{\textit{in-domain evaluations}} across various model sizes. Notably, the compact MiniLMv2 model, with just 23M parameters, surpasses the larger 7B E5-Mistral baseline and 3B cross-encoder monoT5 by achieving an average of 11\% improvement over the baseline on the DL 19/20 datasets. In \textbf{\textit{OOD scenarios}}, \textsc{MuGI} continues to outperform baselines, with gains exceeding 4\% across all models. Specifically, the 23M MiniLMv2 shows that compact bi-encoders can be an effective and robust reranker when equipped with sufficient contextual information. With the \textsc{MuGI} application, it consistently outperforms the larger 7B E5-Mistral model and approaches the performance of the monoT5 3B. However, only modest improvements are noted on the Signal dataset, likely due to the ambiguous queries like \textit{"A stadium for Hughes"} and \textit{"Revilla wants 34 defense witnesses"}, which may not provide sufficient context for LLMs to generate useful pseudo-references.

\subsection{Explore Best Practice of Query Expansion}

\begin{figure*}[th]
    \centering
    \begin{minipage}[t]{.65\textwidth}
        \centering
        \captionsetup{font=small}
        \includegraphics[width=1\linewidth]{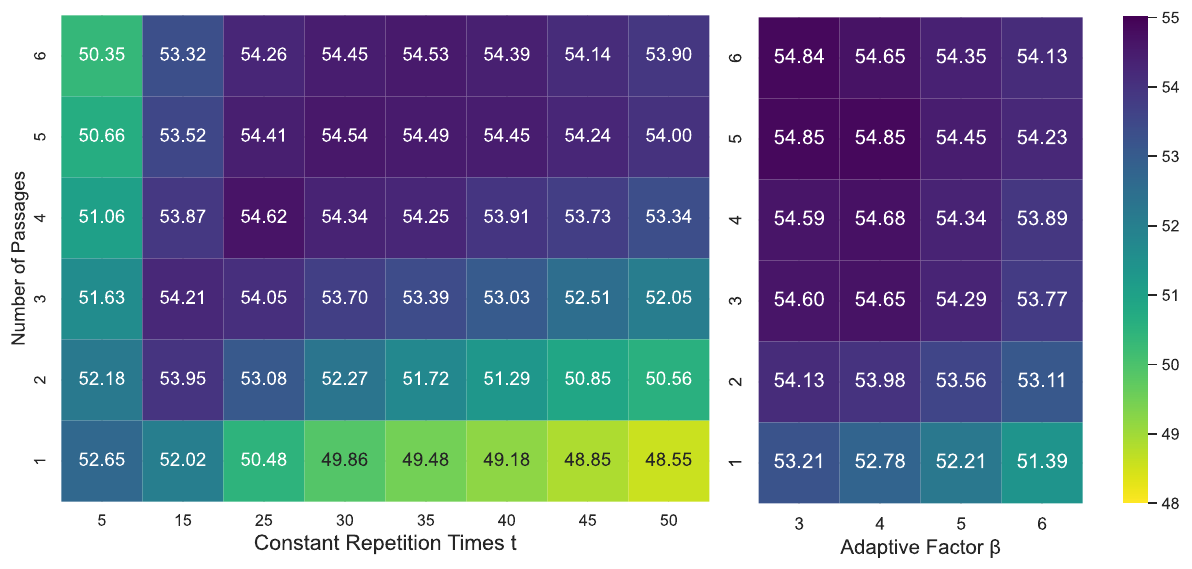}
        \caption{\textbf{BM25 + MuGI} Reweighting Strategy Results (nDCG@10) on average scores of TREC DL + BEIR. The left panel illustrates the constant repetition of the query, while the right panel displays our adaptive reweighting strategy with various $\beta$ values. The Y-axis represents the number of pseudo-references used.
}
        \label{fig:reweight}
    \end{minipage}%
    \hspace{2mm} 
    \begin{minipage}[t]{.3\textwidth}

        \centering
        \captionsetup{font=small}
        \includegraphics[width=\linewidth]{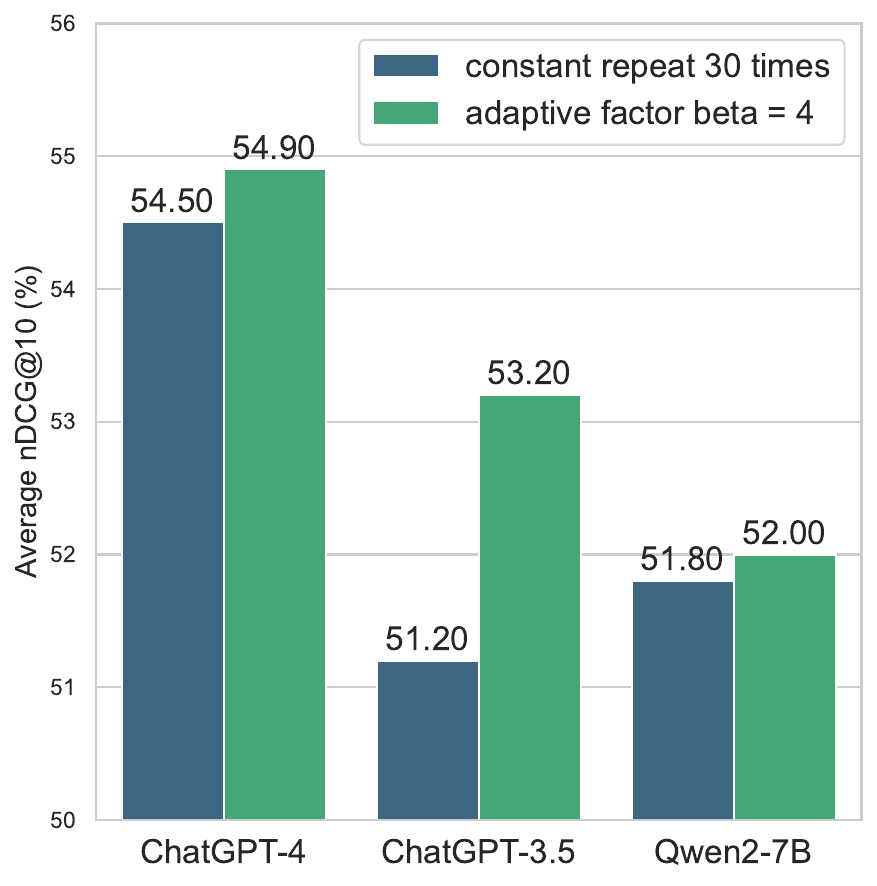}
        \caption{\textbf{BM25 + MuGI over various LLMs with different reweight strategy} with 5 References Results (nDCG@10) .  }
        \label{fig:robust}
    \end{minipage}
    
\end{figure*}

\paragraph{Explore Reweight for BM25.}

Reweighting queries and pseudo references is crucial for optimizing the sensitive BM25 algorithm. To explore reweighting strategy, we experiment with pseudo-references generated by GPT-4 under the \textsc{MuGI} framework in two settings: 1) constant repetition of query for $t$ times and 2) our proposed adaptive reweighting with reweight factor $\beta$.

Fig. \ref{fig:reweight} displays the average scores for the TREC DL+BEIR dataset. It shows that constant repetition, overlooking variations in pseudo-reference lengths, is not a general solution. For instance, repeating a query five times, as suggested in \cite{wang2023query2doc}, is only optimal with a single pseudo-reference. When dealing with multiple pseudo-references, a higher repetition $t$ is needed to achieve better performance. The main drawback of this method is its failure to adjust to the different lengths of references, resulting in inconsistent performances when using a fixed repetition rate $t$. Moreover, finding an effective $t$ for multiple references requires considerable effort.
\begin{table}[th]
    \centering
 
    \resizebox{\linewidth}{!}{
        \begin{tabular}{l|c|c|c|c}
            \toprule
            \textbf{Model} & \textbf{Params} & \textbf{Method} & \textbf{TREC DL} & \textbf{BEIR} \\
            \midrule
            \multirow{2}{*}{MiniLM-L6-v2} & \multirow{2}{*}{23M} 
            & concat & 71.6 & 48.1 \\
            & & feature pool & \textbf{73.6} (\textcolor{darkgreen}{+2.0}) & \textbf{49.5} (\textcolor{darkgreen}{+1.4}) \\
            \midrule
            \multirow{2}{*}{AMB-v2} & \multirow{2}{*}{110M} 
            & concat & 75.7 & 50.3 \\
            & & feature pool & \textbf{76.1} (\textcolor{darkgreen}{+0.4}) & \textbf{50.9} (\textcolor{darkgreen}{+0.6}) \\
            \midrule
            \multirow{2}{*}{Ember-v1} & \multirow{2}{*}{335M} 
            & concat & 74.0 & 52.0 \\
            & & feature pool & \textbf{74.8} (\textcolor{darkgreen}{+0.8}) & \textbf{52.5} (\textcolor{darkgreen}{+0.5}) \\
            \midrule
            \multirow{2}{*}{BGE-Large-EN-v1.5} & \multirow{2}{*}{335M} 
            & concat & 74.3 & 51.2 \\
            & & feature pool & \textbf{74.8} (\textcolor{darkgreen}{+0.5}) & \textbf{51.6} (\textcolor{darkgreen}{+0.4}) \\
            \midrule
            \multirow{2}{*}{E5-Mistral-instruct} & \multirow{2}{*}{7B} 
            & concat & 74.0 & 53.0 \\
            & & feature pool & \textbf{75.8} (\textcolor{darkgreen}{+1.8}) & \textbf{53.6} (\textcolor{darkgreen}{+0.6}) \\
            \bottomrule
        \end{tabular}
    }
           \captionsetup{font=small}
    \caption{Evaluation Results (nDCG@10 \%) of different sized Models with distinct integration approach.}
    \label{tab:integration}
\end{table}
Our adaptive reweighting strategy dynamically adjusts to query and reference lengths, optimizing word frequency in the enhanced query which is important for lexical-based retrievers. This method effectively manages various numbers of references without needing repeated trials to find the optimal repetition ratio. For instance, setting $\beta=4$, as shown in \cref{fig:reweight} (right), consistently yields strong performance across different numbers of reference passages, eliminating the need for a grid search to adjust repetition times $t$ as the number of passages varies.

The adaptive reweighting also maintains robust performance across diverse LLMs, accommodating their varied output lengths. We applied the optimal configuration from \cref{fig:reweight}, which includes a constant repetition of $t=30$ and $\beta=4$ across 5 passages from various LLMs. As demonstrated in \cref{fig:robust}, this adaptive approach consistently outperforms constant repetition. For instance, against ChatGPT-3.5, known for its shorter responses, our method effectively compensates for the inadequate query weighting of constant repetition, ensuring strong performance across different models.

\paragraph{Explore Integration Approach for Dense Retriever.}

We explored two methods for integrating information from pseudo-references with a query: \textbf{\textit{concatenation}} in the input space and \textbf{\textit{feature pooling}} in the feature space. Our experiments with \textsc{MuGI}, conducted without calibration, indicate that feature pooling consistently outperforms simple concatenation, as detailed in \cref{tab:integration}. The primary drawback of concatenation is truncation; given most models' input length limit of 512 tokens, only 1-2 pseudo-references can be accommodated, limiting the utilization of multiple references.

Additionally, concatenation increases computational costs due to its quadratic complexity. For example, with $n$ references each of average length $d$, the complexity of concatenation is $O(n^2d^2)$, compared to $O(nd^2)$ for feature pooling.

\paragraph{Effect of Calibration}

\begin{figure}
    \centering
    \includegraphics[width=1\linewidth]{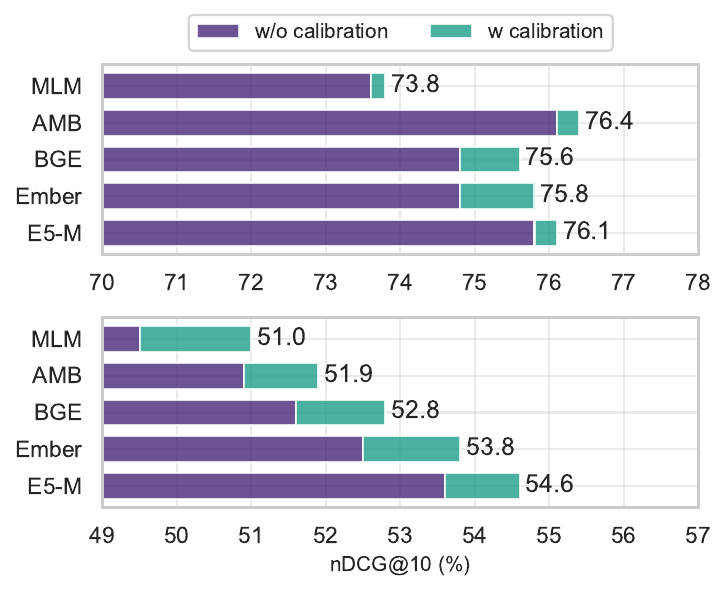}
    \captionsetup{font=small}
    \caption{Calibration Ablation $\alpha=0.2$ (nDCG@10). (Top) In domain TREC DL evaluation; (Bottom) BEIR OOD evaluation. E5-M is E5-Mistral-instruct, BGE is BGE-Large-EN-v1.5, MLM is all-MiniLM-L6-v2, Ember is Ember-v1.}
    \label{fig:calibration}
\end{figure}

The new calibration method leverages hard negative reference feedback to refine query embeddings in the feature space. As shown in \cref{fig:calibration}, it consistently boosts performance in both in-domain and OOD evaluations, with notable gains in OOD scenarios.

%% file: tables/retrieval.tex
\begin{table*}[th]
\centering
\small
\setlength\tabcolsep{2pt}
\resizebox{\linewidth}{!}{

\begin{tabular}{lc c cccccccc c}

\toprule
\multirow{2}{*}{\bf Methods} & \multicolumn{2}{c}{\bf In-domain} & \multicolumn{9}{c}{\bf Out-of-Domain} \\
\cmidrule(rr){2-3}\cmidrule(rrrrrrrrr){4-12}
&DL19 & DL20 & Covid &  NFCorpus &  Touche &  DBPedia & SciFact &  Signal & News &  Robust04 & BEIR (Avg) \\ \midrule

\multicolumn{12}{c}{\textit{\textbf{Supervised}}}\\

ANCE \dag
& 64.5 & 64.6 & 65.4 & 23.7 & 24.0  & 28.1 & 50.7 &24.9 & 38.2 & 39.2 & 36.8\\

DPR\dag 
&62.2 & 65.3 & 33.2 & 18.9 & 13.1 & 26.3 & 31.8 & 15.5 & 16.1 & 25.2 & 22.5\\
DocT5query\dag & 64.2 & - & 71.3 & 32.8 & 34.7 & 33.1 & 67.5 & 30.7 & 42.0 & 43.7 & 44.5 \\
\midrule
\multicolumn{12}{c}{\textit{\textbf{Unsupervised }}} \\

$\text{HyDE\dag}_{\text{Contriever}}$
& 61.3 & 57.9 & 59.3 & - & - & 36.8 & 69.1 & - & 44.0 & - & - \\

\rowcolor{lightgray}BM25
& 50.6 & 48.0 & 59.5 & 30.8 & 44.2 & 31.8 & 67.9 & 33.1 & 39.5 & 40.7 & 43.4
\\

\quad + \textit{RM3}\dag
& 52.2 & 47.4 & - & -& -& - & -& - & - & - & - \\ 

\quad + \textit{query2doc}\dag
& 66.2 & 62.9 & 72.2 & 34.9 & 39.8 & 37.0 & 68.6 & - & -&-&- \\

\quad + \textit{LameR }\dag
& 67.1 & 62.7 & 72.5 & -& -& 38.7 & 73.5 & - & 49.9 & - & - \\ 

\quad + \textit{MuGI (ChatGPT-3.5)}
& 70.4 & 63.9 & 69.7 & 36.0& 46.3& 41.2 & 72.0 & 35.8 & 46.6 & 49.7 & 49.6 \\ 

\quad + \textit{MuGI (ChatGPT-4)}
& $\textbf{70.4}^\text{+19.8}$ & ${\textbf{64.4}}^{\text{+16.4}}$ & $\textbf{72.9}^{\text{+13.4}}$ & $\textbf{37.4}^{\text{+6.6}}$ & $\textbf{46.1}^{\text{+1.9}}$ & $\textbf{42.7}^{\text{+10.9}} $& $\textbf{74.0}^{\text{+6.1}}$ & $\textbf{36.0}^{\text{+2.9}}$ & $\textbf{50.0}^{\text{+10.5}}$ & $\textbf{49.2}^{\text{+8.5}}$ &  $\text{\textbf{51.0}}^{\text{+7.6}}$ \\



\bottomrule
\end{tabular}}


\caption{\textbf{Retrieval Results (nDCG@10).} Best performing are marked bold. \dag represents cited results. 
}
\label{table:retrieval}
\end{table*}

%% file: tables/rerank.tex
\begin{table*}[!ht]
\centering
\small
\setlength\tabcolsep{2pt}
\resizebox{\textwidth}{!}{
\begin{tabular}{lc cc  cccccccc  c }

\toprule
\multirow{2}{*}{\bf Methods} & \multirow{2}{*}{\#Param} & \multicolumn{2}{c}{\bf In-domain} & \multicolumn{9}{c}{\bf Out-of-Domain} \\
\cmidrule(rr){3-4}\cmidrule(rrrrrrrrr){5-13}
&&DL19 & DL20 & Covid &  NFCorpus &  Touche &  DBPedia & SciFact &  Signal & News &  Robust04 & BEIR(Avg) \\ \midrule

\multicolumn{13}{c}{\textit{\textbf{Dense Cross-Encoder + BM25 re-rank}}}
 \\


monoT5 \dag&220M
& 71.5 & 67.0 & 78.3 & 37.4 & 30.8 & 42.4 & 73.4 & 31.7 & 46.8 & 51.7 & 49.1
\\

monoT5\dag &3B
& 71.8 & 68.9 & 80.7 & 39.0 & 32.4 & 44.5 &  76.6 & 32.6 & 48.5 & 56.7 & 51.4
\\

RankLLaMA\ddag &7B
& 75.6 & 77.4 & 85.2 & 30.3 & 40.1 & 48.3 &  73.2 & - & - & - & -
\\

Cohere Rerankv2 \dag& API
& 73.2 & 67.1 & 81.8 & 36.4 & 32.5 & 42.5 & 74.4 & 29.6 & 47.6 & 50.8 & 49.5
\\
\midrule
\multicolumn{13}{c}{\textit{\textbf{Dense Bi-Encoder + Query Expansion}}}
\\

\rowcolor{lightgray}all-MiniLM-L6-v2 &23M &&&&&&&&&&&\\
\quad + \textit{BM25 re-rank } & & 64.2 & 60.8 & 73.6 & 30.8 & 26.2 & 35.9 & 67.6 & 28.8 & \textbf{52.0} & 51.4 & 45.8\\

\quad + \textit{query2doc } & & 72.1 & 69.3 & - &- & - & - & - & - & - &- & -\\

\quad + \textit{MuGI Pipeline}
& & ${\textbf{75.4}}^{\text{+11.2}}$ & $\textbf{\text{72.1}}^{\text{+11.3}}$ & ${\textbf{81.0}}^{\text{+7.4}}$ & $\textbf{37.6}^{\text{+6.8}}$ & $\textbf{34.1}^{\text{+7.9}}$ & $\textbf{45.5}^{\text{+9.6}}$ & $\textbf{74.8}^{\text{+7.2}}$ 
& $\textbf{28.9}^{\text{+0.1}}$& 
$51.4$ 
& $\textbf{55.2}^{\text{+3.8}}$ & $\textbf{51.0}^{\text{+5.2}}$\\

\rowcolor{lightgray}AMB-v2 &110M &&&&&&&&&&&\\
\quad + \textit{BM25 re-rank } & & 68.3 & 64.2 & 75.8 & 34.3 & 28.9 & 37.6 & 68.0 & 29.8 & 52.0 & 51.4 & 47.2\\
\quad + \textit{query2doc } & & 74.5 & 73.3 & - &- & - & - & - & - & - &- & -\\

\quad + \textit{MuGI Pipeline}
& & ${\textbf{77.6}}^{\text{+9.3}}$ &
$\textbf{\text{75.1}}^{\text{+10.9}}$ &
${\textbf{82.6}}^{\text{+6.8}}$ &
$\textbf{39.2}^{\text{+4.9}}$ & 
 $\textbf{30.3}^{\text{+1.4}}$ &
$\textbf{45.7}^{\text{+7.8}}$ &
$\textbf{73.9}^{\text{+5.9}}$ & 
$\textbf{30.2}^{\text{+0.4}}$ & 
$\textbf{55.2}^{\text{+3.2}}$ &
$\textbf{58.2}^{\text{+6.8}}$ &
$\textbf{51.9}^{\text{+4.7}}$ \\

\rowcolor{lightgray}BGE-Large-EN-v1.5 &335M&&&&&&&&&&& \\
\quad + \textit{BM25 re-rank } & &	70.7& 64.9&	79.6& 35.4 &	29.8&	41.7&	74.1&	28.8&	49.4&	49.2&	48.5 \\
 \quad + \textit{query2doc } & & 71.7 & 70.2 & - &- & - & - & - & - & - &- & -\\

\quad + \textit{MuGI Pipeline}
&  &	$\textbf{78.1}^{\text{+7.4}}$ & $\textbf{73.0}^{\text{+8.1}}$ & $\textbf{84.3}^{\text{+4.7}}$ & $\textbf{40.9}^{\text{+5.5}}$ & $\textbf{33.1}^{\text{+3.3}}$ & $\textbf{47.0}^{\text{+5.3}}$ & $\textbf{77.8}^{\text{+3.7}}$ & $\textbf{30.0}^{\text{+1.2}}$ & $\textbf{53.6}^{\text{+4.2}}$ & $\textbf{55.3}^{\text{+6.1}}$ & $\textbf{52.8}^{\text{+4.3}}$ \\

\rowcolor{lightgray}Ember-v1 &335M  &&&&&&&&&&& \\
\quad + \textit{BM25 re-rank } & & 71.3 & 64.5 & 80.3 & 35.8 & 31.6 & 41.7& 75.2& 32.0 & 48.8 & 51.3 & 49.6
\\

\quad + \textit{query2doc } & & 71.2 & 69.9 & - &- & - & - & - & - & - &- & -\\

\quad + \textit{MuGI Pipeline}
& &	$\textbf{78.3}^{\text{+7.0}}$ & $\textbf{73.3}^{\text{+8.8}}$ & $\textbf{84.7}^{\text{+4.4}}$ & $\textbf{40.7}^{\text{+4.9}}$ & $\textbf{35.8}^{\text{+4.2}}$ & $\textbf{46.6}^{\text{+4.9}}$ & $\textbf{78.3}^{\text{+3.1}}$ & $\textbf{32.6}^{\text{+0.6}}$ & $\textbf{54.0}^{\text{+5.2}}$ & $\textbf{58.0}^{\text{+6.7}}$ & $\textbf{53.8}^{\text{+4.2}}$ \\

\rowcolor{lightgray}E5-Mistral-instruct & 7B &&&&&&&&&&& \\
\quad + \textit{BM25 re-rank } & &	70.0& 66.7&	81.4&	36.0&	29.5&	42.4&	75.8&	32.8&	53.0&	52.8&	50.5 \\ 

\quad + \textit{query2doc } & & 71.3 & 71.6 & - &- & - & - & - & - & - &- & -\\

\quad + \textit{MuGI Pipeline}
& &	$\textbf{77.3}^{\text{+7.3}}$ & $\textbf{74.9}^{\text{+8.2}}$ & $\textbf{85.6}^{\text{+4.2}}$ & $\textbf{41.3}^{\text{+5.3}}$ & $\textbf{35.8}^{\text{+6.3}}$ & $\textbf{47.3}^{\text{+5.1}}$ &
$\textbf{77.9}^{\text{+2.1}}$  &  $\textbf{34.5}^{\text{+2.4}}$
&  $\textbf{55.1}^{\text{+2.1}}$
&  $\textbf{59.2}^{\text{+7.6}}$
&  $\textbf{54.6}^{\text{+4.1}}$

\\

\bottomrule
\end{tabular}
}
 \captionsetup{font=small}

\caption{\textbf{Re-ranking Results (nDCG@10) on TREC and BEIR.} Best performing are marked bold. MuGI pipeline suggests application of MuGI on both sparse retrieval and dense retrieval as shown in \cref{fig:method}. \ddag RankLLaMA rerank based on top 200 references from RepLLaMA.}
\label{tab:rerank}
\end{table*}

%% file: chapter/analysis.tex
\section{Analysis}

\subsection{Ablation Study}
In this section, we analyze and conduct ablation studies on MuGI using the all-MiniLM-L6-v2 and references generated by GPT-4 within the full MuGI pipeline settings described in \cref{sec:detail}.

\begin{figure}
    \centering
    \includegraphics[width=\linewidth]{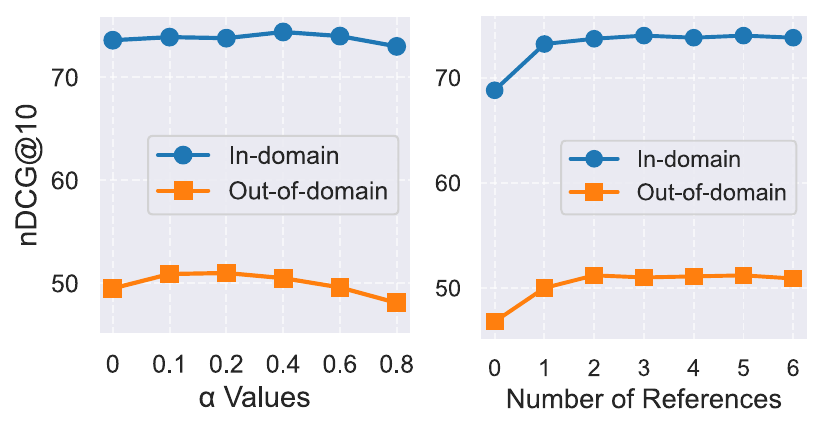}
    \captionsetup{font=small}
    \caption{(Left) Ablation on $\alpha$ in calibration. (Right) Ablation on the number of generated pseudo references for MiniLM-L6. Zero references represent directly rerank on top100 BM25.}
    \label{fig:ablation}
 
\end{figure}

\paragraph{Impact of Number of Pseudo References}
The fundamental premise of MuGI is that multiple references generated by LLMs can provide contextual information and key words or relevant patterns that enhance queries. Consequently, the critical factor in our framework is the number of pseudo references used. Both sparse retrieval (\cref{fig:reweight}) and dense retrieval (\cref{fig:ablation}) show performance improvements with an increasing number of references; however, gains plateau at five. This suggests that the language model’s capability to generate key terms reaches its limit at this point.

\paragraph{Impact of $\alpha$ in Calibration}
Figure \ref{fig:ablation} indicates that the calibration parameter $\alpha$ offers benefits, with performance peaking at 0.2 and then declining. The impact of calibration is more pronounced in OOD evaluations, in line with observations from \cref{fig:calibration}.

\paragraph{Impact of Large Language Models}
We assessed the MuGI framework across various LLMs, specifically examining BM25 (\cref{fig:robust}) and dense retrievers (\cref{tab:llms}). Our findings highlight that: 1) stronger LLMs consistently produce better results; 2) GPT models outperform others, likely due to their expansive knowledge bases. Although GPT-3.5 generally underperforms relative to Qwen2-7B on most open leaderboards \cite{lmsys, Open-LLM-Leaderboard-Report-2023}, it demonstrated superior performance in our tests, possibly due to its more robust generalization capabilities.

\begin{table}[]
    \centering
    \resizebox{\linewidth}{!}{
    \begin{tabular}{lcccccccccc}

     \toprule
        & Qwen2-7B & Qwen2-72B & GPT3.5 & GPT4& GPT4o \\ \midrule
        In-domain & 68.3 & 73.1 & 73.1 & 73.8 & 73.4 \\
        OOD & 47.8 & 50.3 & 50.5 & 51.0 & 50.2 \\ \bottomrule
    \end{tabular}}
     \captionsetup{font=small}
    \caption{nDCG@10 Results Using MiniLM with References from Diverse LLMs.}
    \label{tab:llms}

\end{table}
\subsection{Mechanism behind Query Expansion}

\begin{figure}
    \centering
    \includegraphics[width=\linewidth]{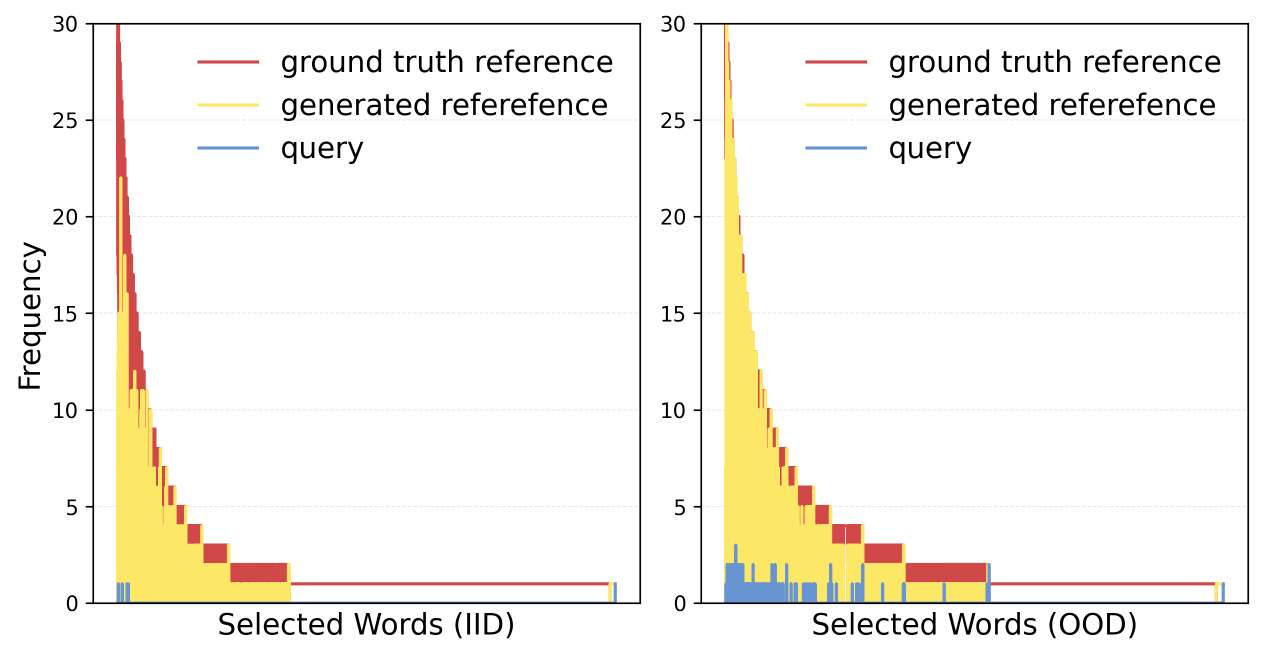}
     \captionsetup{font=small}
    \caption{Key Words Overlap Distribution.}
    \label{fig:tfidf}

\end{figure}

We investigate how generated pseudo references improve IR through key vocabulary overlap. We compile "Ground Truth References (GT)" from passages rated relevance level 3 and aggregate "pseudo references (PSE)" for each query from TREC DL. By identifying the top 10 highest Inverse Document Frequency (IDF) words in these references, we compare the frequency of key vocabularies in the query, GT, and PSE. As illustrated in \cref{fig:tfidf}, there is a substantial overlap in the frequency of key vocabularies between GT (red region) and PSE (yellow region) in both IID and OOD scenarios, surpassing that between the original query (blue region) and GT (red region). This demonstrates that pseudo references significantly enhance retrieval by incorporating crucial key vocabularies or patterns that target specific passages.

%% file: chapter/conclusion.tex
\section{Conclusion}

This paper explores best practices for implementing query expansion methods in information retrieval with large language models. We present the Multi-Text Generation Integration (MuGI) framework, a technique that markedly improves information retrieval by integrating the query with multiple generated passages through an adaptive reweighting strategy, feature pooling, and query calibration. Our empirical findings show that MuGI significantly enhances the performance of both sparse and dense retrieval models. We offer a comprehensive discussion of best practices, informed by thorough experimentation.